\documentclass[pdftex,dvipsnames,usenames,aps,superscriptaddress,twocolumn,amsmath,showkeys,prd]{revtex4-1}

\usepackage{hyperref}
\usepackage{float}
\usepackage{dcolumn}
\usepackage{graphicx}
\usepackage{color}
\usepackage{amssymb}
\usepackage{bold-extra}
\usepackage{subcaption}

\def\VYPx#1#2#3{{\bf #1}, #3 (#2)}  
\newcommand{\VYP}[4][]{%
  \ifx\\#1\\
    \VYPx{#2}{#3}{#4}
  \else
    \href{http://dx.doi.org/#1}{\VYPx{#2}{#3}{#4}}
  \fi
}

\def\etal{ {\it et al.}}
\def\Ttl#1{``{\it #1}'',}

\def\PRD#1#2#3{\href{http://link.aps.org/doi/10.1103/PhysRevD.#1.#3}{Phys.~Rev.~D~\VYP{#1}{#2}{#3}}}

\newcommand{\AeA}[4][]{Astron.~\& Astrophys.\ \VYP[#1]{#2}{#3}{#4}}

\newcommand{\ApP}[4][]{Astropart.~Phys.\ \VYP[#1]{#2}{#3}{#4}}

\newcommand{\IJMPA}[4][]{Int.~J.~Mod.~Phys.~A\ \VYP[#1]{#2}{#3}{#4}}

\newcommand{\JCAP}[4][]{J.~Cosm.~and~Astrop.~Phys.\ \VYP[#1]{#2}{#3}{#4}}

\newcommand{\NIM}[4][]{Nucl.~Instr.~and Meth.~A\ \VYP[#1]{#2}{#3}{#4}}

\newcommand{\PRep}[4][]{Phys.~Rep.\ \VYP[#1]{#2}{#3}{#4}}

\def\arXiv#1{\href{http://arxiv.org/abs/arXiv:#1}{arXiv:{#1}}}
\def\nucl-th#1{\href{http://arxiv.org/abs/nucl-th/#1}{nucl-th/{#1}}}

\renewcommand{\Ttl}[1]{}

\def\figref#1{Fig.~\ref{fig:#1}}
\def\figlab#1{\label{fig:#1}}  
\def\eqref#1{Eq.~(\ref{eq:#1})}

\def\eqlab#1{\label{eq:#1}}

\newcommand*{\appref}[1]{Appendix~\ref{app:#1}}
\newcommand*{\applab}[1]{\label{app:#1}}
\newcommand{\Omit}[1]{}

\def\v{\vec{v}}
\def\B{\vec{B}}
\def\vB{{\vec{v}\times\vec{B}}}
\def\vvB{{\vec{v}\times\vec{v}\times\vec{B}}}
\newcommand{\Xmax}{$X_{\rm max}$}
\newcommand{\tret}{t_{\rm ret}}

\def\KVI{University of Groningen, KVI Center for Advanced Radiation Technology, 9747 AA Groningen, The Netherlands}
\def\AIVUB{Astrophysical Institute, Vrije Universiteit Brussel, Pleinlaan 2, 1050 Brussels, Belgium}
\def\VUB{Vrije Universiteit Brussel, Dienst ELEM, IIHE, Pleinlaan 2, 1050 Brussels, Belgium}
\def\NIKHEF{Nikhef, Science Park Amsterdam, 1098 XG Amsterdam, The Netherlands}
\def\IMAPP{Department of Astrophysics/IMAPP, Radboud University, P.O. Box 9010, 6500 GL Nijmegen, The Netherlands}
\def\ASTRON{Netherlands Institute for Radio Astronomy (ASTRON), 7990 AA Dwingeloo, The Netherlands}
\def\MPIB{Max-Planck-Institut f\"{u}r Radioastronomie, 53121 Bonn, Germany}
\def\UCI{Physics and Astronomy, University of California, Irvine, CA 92697-4575,U.S.A}
\def\Vax{Department of Physics and Electrical Engineering, Linn\'{e}universitetet, 35195 V\"{a}xj\"{o}, Sweden}
\def\Prince{Department of Astrophysical Sciences, Princeton University, Princeton, NJ 08544, USA}
\begin{document}

\title{Measurement of the circular polarization in radio emission from extensive air showers confirms emission mechanisms.
}

\author{O.~Scholten} \email[]{O.Scholten@rug.nl} \affiliation{\KVI}   \affiliation{\VUB}
\author{T.~N.~G.~Trinh} \email[]{t.n.g.trinh@rug.nl}  \affiliation{\KVI}
\author{A.~Bonardi} \affiliation{\IMAPP}
\author{S.~Buitink} \affiliation{\AIVUB}
\author{P.~Correa} \affiliation{\VUB}
\author{A.~Corstanje} \affiliation{\IMAPP}
\author{Q.~Dorosti Hasankiadeh} \affiliation{\KVI}
\author{H.~Falcke}  \affiliation{\IMAPP} \affiliation{\NIKHEF} \affiliation{\ASTRON} \affiliation{\MPIB}
\author{J.~R.~H\"orandel}  \affiliation{\IMAPP} \affiliation{\NIKHEF}
\author{P.~Mitra} \affiliation{\AIVUB}
\author{K.~Mulrey} \affiliation{\AIVUB}
\author{A.~Nelles}  \affiliation{\IMAPP} \affiliation{\UCI}
\author{J.~P.~Rachen} \affiliation{\IMAPP}
\author{L.~Rossetto}  \affiliation{\IMAPP}
\author{P.~Schellart} \affiliation{\IMAPP} \affiliation{\Prince}
\author{S.~Thoudam} \affiliation{\IMAPP} \affiliation{\Vax}
\author{S.~ter Veen} \affiliation{\IMAPP} \affiliation{\ASTRON}
\author{K.D.~de~Vries} \affiliation{\VUB}
\author{T.~Winchen} \affiliation{\AIVUB}

\date{\today}

\begin{abstract}
We report here on a novel analysis of the complete set of four Stokes parameters that uniquely determine the linear and/or circular polarization of the radio signal for an extensive air shower.
The observed dependency of the circular polarization on azimuth angle and distance to the shower axis is a clear signature of the interfering contributions from two different radiation mechanisms, a main contribution due to a geomagnetically-induced transverse current and a secondary component due to the build-up of excess charge at the shower front. The data,  as measured at LOFAR, agree very well with a calculation from first principles. This opens the possibility to use circular polarization as an investigative tool in the analysis of air shower structure, such as for the determination of atmospheric electric fields.
\end{abstract}


\maketitle

\section{Introduction}

Detecting the radio emission from extensive air showers (EASs) as induced by energetic cosmic rays has shown to be a very sensitive way to determine shower properties, such as energy~\cite{LOFAR-E,AERA-E} and \Xmax, the atmospheric (slant) depth where the number of air-shower particles reaches a maximum~\cite{Stijn14,Hue16}. These shower properties are used in turn to infer the nature of the primary cosmic ray, in particular its  mass~\cite{Lopes-Xm,Bui16}.
To do so, detailed models have been developed that calculate the radio emission from the EAS based on the motion of individual electrons and positrons in the air shower. Two such microscopic models are CoREAS~\cite{COREAS} and ZHAireS~\cite{ZHairS}.
Their close agreement with detailed radio-intensity footprints, as have been measured with LOFAR (LOw-Frequency ARray)~\cite{Sche13,Nel14a} and other radio antenna arrays~\cite{CODALEMA}, shows that the microscopic approach can successfully reproduce the features of the radio emission~\cite{Hue16}.
The signal has a dominant linearly-polarized component along the direction of the Lorentz force, $\hat{e}_{\vB} \propto \hat{e}_{\v}\times\hat{e}_{\B}$, due to the induced transverse current in the shower front. Here the shower direction is given by $\hat{e}_{\v}$ while $\hat{e}_{\B}$ denotes the direction of the geomagnetic field.
A secondary contribution, also known as Askaryan radiation~\cite{Askaryan}, is due to the build-up of excess negative charge in the shower front~\cite{Krijn10}. This Askaryan radiation is radially polarized and thus leads to the prediction that the deviation from the main polarization direction depends on the viewing angle, in good agreement with observations~\cite{Mar2011,Scho12,PAO14,Sche14,Bel15}. The microscopic approaches to simulate the radio emission make no explicit assumption about different emission mechanisms, instead their predictions arise from first-principle calculations. Describing the emission based on emission mechanisms has shown, however, to be a helpful tool in understanding event properties.

From the work of Ref.~\cite{Schoo08} it is known that the radio pulse has also a certain amount of circular polarization. A very appropriate way to express the complete polarization of radio pulse is through the use of Stokes parameters~\cite{Sche14} and these have more recently also been used in the analysis of ANITA data~\cite{Gor16} of cosmic-ray events.
Despite surprisingly high measured fractions of circular polarization in reflected signals reported by the ANITA collaboration, CoREAS simulations and other models \cite{Hue16} predict the percentage of circular polarization in direct air shower signals to be small.
In this work we report on the measurement at LOFAR of the circular-polarization footprint of an EAS and present a novel analysis in terms of a phase (time) delay which is an earmark of an important difference between the main, the transverse current, and the secondary, charge excess, emission processes.
Due to a different dependence of these emission on the viewing angle of the electric currents in the shower~\cite{Krijn10,Vri13}, as shown later, a slight, order 1~ns, time-shift between the pulses emitted by the two emission mechanisms is created, resulting in a rotation of the polarization vector over the duration of the pulse. This circular polarization thus constitutes an accurate measurement of the difference in the arrival times of the two components.

\section{LOFAR}

LOFAR~\cite{Haa13} is a digital radio telescope with Low Band Antennas (LBAs, 10 - 90~MHz band) and High Band Antennas (HBAs, 110 - 240~MHz band). Each LBA consists of two inverted V-shaped dipoles labeled X and Y.
Radio emission from cosmic rays has been measured with LBAs as well as HBAs~\cite{Sche13,Nel14a}; here we focus on LBA measurements, as they record stronger air shower signals and their hardware set-up provides a cleaner way to disentangle polarization properties from single antenna measurements than the HBAs.
A detailed description of the offline analysis of the data can be found in~\cite{Sche13,Stijn14,Sche14}. However, we will briefly review the essential steps in reconstructing the full electric-field vector. For every detected air shower, the time-dependent voltages as measured with the X and Y dipoles of the LBAs are available.
Usually, 2.1 ms of data, sampled at 200~Msamples/s are recorded per antenna. Each dipole trace shows a short pulse of a length that is determined by the intrinsic length of pulse in combination with the limited band width of the antenna and the filter amplifier of the LOFAR system. The frequency spectrum of the pulses is usually strongest at the lowest frequencies and especially the high-frequency component changes strongly as function of distance to the shower axis~\cite{Nel14a}, which influences the intrinsic length of the pulses. The recorded pulses are usually strong and their amplitudes are commonly a factor of 10 or more larger than the average noise amplitude in a single dipole~\cite{Sche14}. The same read-out hardware is used for all signals, so time-shifts artificially introduced between a pair of dipoles are negligible. The antenna response of a pair of dipoles has been simulated for all possible arrival directions and has been shown to be in good agreement with dedicated in-situ measurements~\cite{Nel15}. The complex (i.e.\ time-dependent) responses of each dipole to an incoming wave of a given polarization are tabulated~\cite{Sche13} and are used in an iterative approach to obtain an arrival direction and to unfold the antenna response, as for a cosmic rays the arrival direction is not known a priori. The simulated antenna response allows for a time-dependent reconstruction of the on-sky polarization (i.e.\ two perpendicular components of the transverse electric field) of the incoming signal from the measured voltages without additional assumptions about the content of cross-polarization leakage and projections. Using the reconstructed arrival direction, these on-sky polarizations can be rotated in the direction of the Lorentz force  $\hat{e}_{\vB}$ and the perpendicular component $\hat{e}_{\vvB}$. The thus reconstructed electric field, is the band-width limited original electric field. In this process, all instrumental influences induced by the antenna and the amplifier have been corrected for. Influences induced by the LOFAR system, such as frequency dependent cable delays and additional amplification~\cite{Nel15}, have, however, not been corrected for. As they influence both signal paths of a pair of dipoles equally, they do not induce a change in polarization that would be relevant for this analysis, but might increase the length of the pulse.

For the course of this analysis, the antenna positions known to centimeter accuracy on ground are projected onto the shower plane, defined as the plane perpendicular to the shower, $\hat{e}_{\v}$, and going through the point where the core of the shower hits the ground.

\begin{figure}
  \includegraphics[width=.47\textwidth]{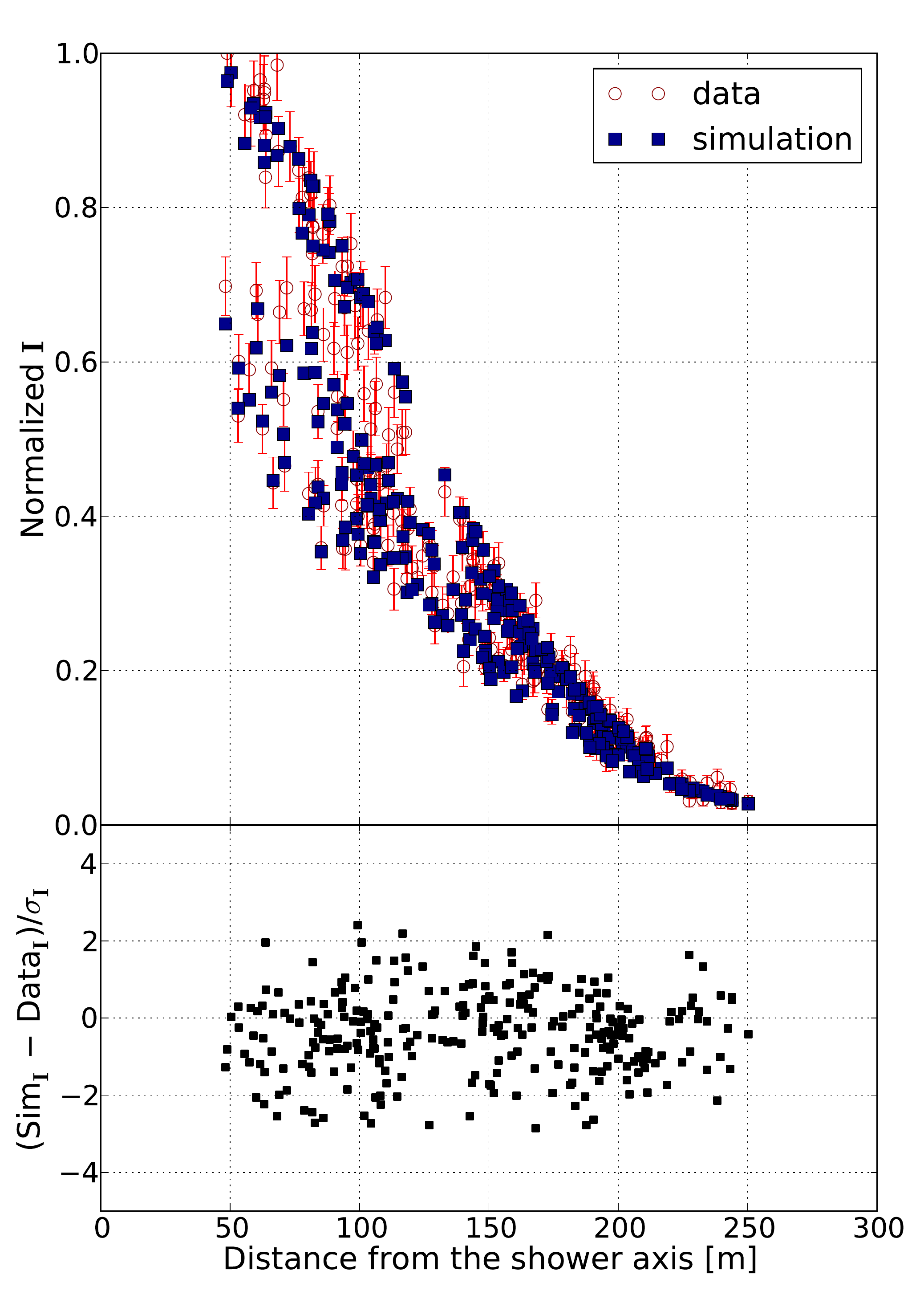}
  \caption{The intensity footprint of the air shower we consider here as recorded with the LOFAR low-band antennas and projected into the shower plane (open red circles) is compared to the results of a CoREAS simulation (filled blue squares). $\sigma$ denotes one standard deviation error.\figlab{StokesI_5}
  } \end{figure}

\begin{figure*}
 \centerline{\includegraphics[width=.99\textwidth]{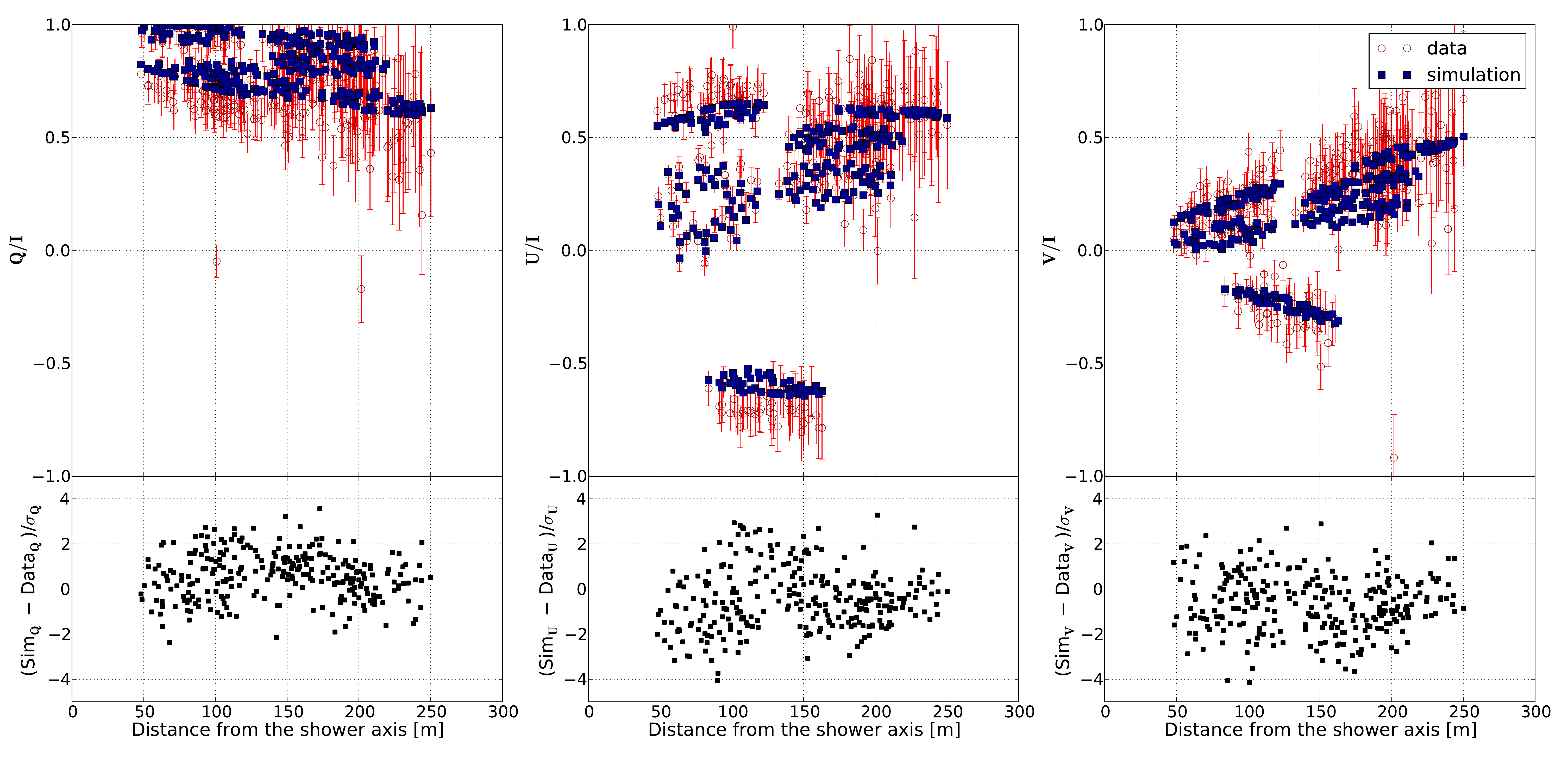}}
  \caption{The set of normalized Stokes parameters that characterize the polarization footprint of a single air shower. Refer to the caption of \figref{StokesI_5} for the meaning of the symbols.
  \figlab{Stokes-ratio}}
\end{figure*}

\section{Data analysis}

The Stokes parameters can be expressed in terms of the complex voltages ${\cal E}_i=E_i + i\hat{E}_i$, where $E_i$ is sample $i$ of the electric field component in either the $\hat{e}_\vB$ or the $\hat{e}_\vvB$ polarization direction and $\hat{E}_i$ its Hilbert transform~\cite{Sche14}, as
\begin{eqnarray}
I&=&{1\over n} \sum_0^{n-1} \left( |{\cal E}|^2_{i,\vB} + |{\cal E}|^2_{i,\vvB} \right) \nonumber\\
Q&=&{1\over n} \sum_0^{n-1} \left( |{\cal E}|^2_{i,\vB} - |{\cal E}|^2_{i,\vvB} \right) \nonumber\\
U +iV&=&{2\over n} \sum_0^{n-1} \left( {\cal E}_{i,\vB} \;  {\cal E}_{i,\vvB}^* \right) \;,
\eqlab{Stokes}
\end{eqnarray}
where one may define in addition $W^2=I^2-( Q^2 + U^2 + V^2)$.
The summations are performed over $n = 5$ samples, of 5~ns each, centered around the pulse maximum,   which is determined by the strongest signal in one of the electric field components. This time scale has been found to contain the entire pulse, also for pulses at larger distance to the shower axis, where the high frequency content of the pulses is reduced and their intrinsic length increased.
The linear-polarization direction of the signal is given by the angle $\psi={1\over 2} \tan^{-1} (U/Q)$ with the $\vB$-axis, while $V/I$ specifies the circular polarization.
For $n=1$ in \eqref{Stokes} one obtains $W=0$, however in general $W\ge 0$ is a measure of the difference in structure of the signal in the two polarization directions, which is likely due to noise (see \appref{W} for a more extensive discussion).  It has been reported earlier \cite{Sche14} that all recorded air shower pulses show a strong fraction of polarization, where the unpolarized component decreases when the signal-to-noise ratio increases. We therefore use a function dependent on the signal-to-noise ratio to calculate the relevant uncertainties. On average the signal-to-noise ratio decreases with increasing distance to the shower axis. Additional uncertainties arise from the uncertainty on the reconstructed arrival direction and the antenna model. Both are propagated accordingly as discussed in \cite{Sche14}.

Qualitatively the presence of circular polarization can be understood as due to the arrival-time difference of the signals in the two polarization directions. The polarization of the field thus rotates over the duration of the pulse. To make this more quantitative we interpret the circular polarization in terms of a time lag between the transverse current and the charge-excess pulses. To do so it is conceptually easiest to consider a signal of fixed frequency $\omega$. We assume a phase difference
\begin{equation}
\eta=\omega \Delta t
\eqlab{eta}
\end{equation}
that corresponds to a delay $\Delta t$ of the radially-polarized charge-excess pulse,
$${\cal E}_C(t)=E_C e^{i\omega t-i\eta} (\cos{\phi}\, \hat{e}_{\vB}+\sin{\phi}\, \hat{e}_{\vvB}) \;,$$
with respect to the transverse-current pulse,
$${\cal E}_T(t)=E_T e^{i\omega t} \hat{e}_{\vB}\;.$$
Here $\phi$ denotes the angular position of the antenna with respect to the $\hat{e}_{\vB}$-direction. Substituting this into \eqref{Stokes} yields
\begin{eqnarray}
I&=& E_T^2 + E_C^2 + 2 \cos\phi \cos\eta E_T E_C \nonumber\\
Q&=& E_T^2 +\cos(2\phi) E_C^2 + 2 \cos\phi \cos\eta E_T E_C  \nonumber\\
U&=& \sin(2\phi) E_C^2 + 2 \sin\phi \cos\eta E_T E_C  \nonumber\\
V&=&   2\sin\phi \sin\eta E_T E_C  \;.
\eqlab{Stokes-anal}
\end{eqnarray}
For $\phi=0$ or $\phi=\pi$ one obtains obtains $U=0=V$ i.e.\ no circular polarization and full linear polarization in the $\hat{e}_{\vB}$-direction.
Extreme values for the circular polarization are reached for $\phi=\pi/2$,  giving $I=E_T^2+E_C^2$, $U=E_T E_C \cos{\eta}$ and $V= E_T E_C \sin{\eta}$ while  $\phi=-\pi/2$ yields the opposite signs for $U$ and $V$. This shows that $U/I$ is a measure for the relative strength of the charge excess, $E_C$, and the transverse current, $E_T$, components~\cite{Krijn10} while $V/U=\tan{\eta}$ measures the phase delay, and thus the time-lag, between the two.

For sake of concreteness we initially focus the present discussion on a single air shower for which the radio signal was detected in six LOFAR stations each consisting out of 48 antennas. For this event the full set of measured Stokes parameters for each LOFAR antenna is shown versus distance with respect to the shower axis in \figref{StokesI_5} and \figref{Stokes-ratio}.
Also shown are the results of the CoREAS simulation,  where the values for \Xmax$=659$~g/cm$^2$, the shower energy $E=6.24\times10^{17}$~eV, zenith angle $\theta=26^\circ$, and core position, have been taken from a fit solely to the intensity footprint~\cite{Stijn14}.
Like in Ref.~\cite{Stijn14} the simulations are performed for a star-shaped pattern of antennas and interpolated to obtain the results at the actual positions of the antennas. The interpolation is accurate to within 10\% at distances used in the present analysis.
The calculated values for $W/I$ (not shown) are small, less than a few percent.
Due to noise the measured values for $W/I$ are generally larger  {than predicted without noise, with a bias} increasing from less than a percent at small distances to less than 10\% at larger distances.
The values of the Stokes parameters depend not only on distance with respect to the shower axis but also on the azimuthal position of the antenna, see \eqref{Stokes-anal}. Thus, when plotted versus distance only, one observes a considerable scatter in the data points, reflecting the layout of the antenna stations.

The angular dependence of the circular polarization is most clearly seen in \figref{Stokes-V} where the footprint of the Stokes parameter $V$ is shown as obtained from the simulation and data.  As expected, see \eqref{Stokes-anal}, $\hat{e}_{\vB}$ is the axis of anti-symmetry, where $V$ changes sign along  $\hat{e}_{\vvB}$ to -$\hat{e}_{\vvB}$.

\begin{figure}
  \includegraphics[width=.49\textwidth]{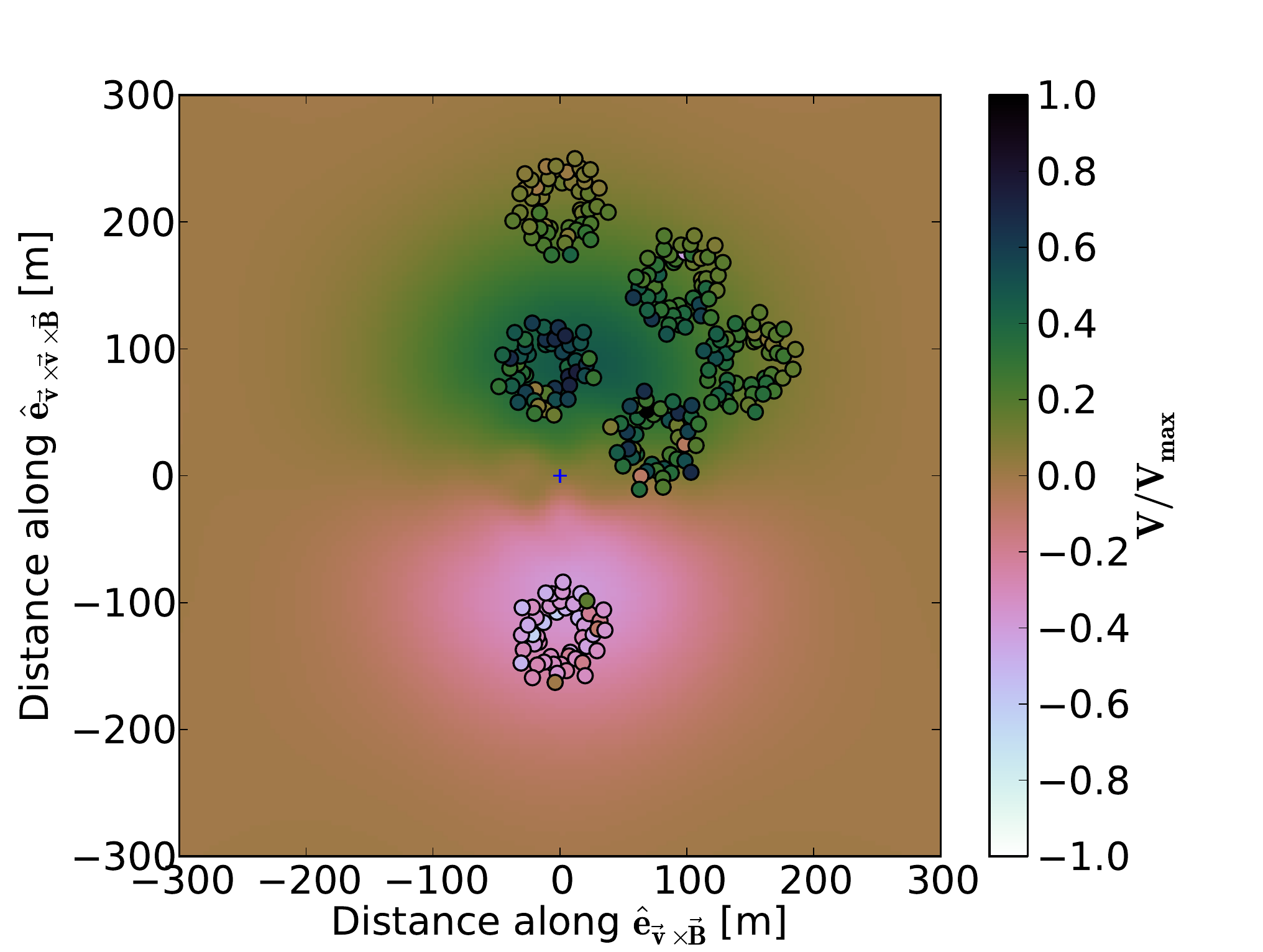}
  \caption{The footprint of the value of the Stokes $V$-parameter for a measured air shower. The background color shows the results of the CoREAS simulation while the coloring in the small circles presents the data. This is the same data as shown in \figref{Stokes-ratio} (right most panel), however not normalized by $I$ but by the maximum of V. At close distances to the shower axis (cross) the predicted values for V suffer from numerical instability in the simulation.
  \figlab{Stokes-V}}
\end{figure}

In analyzing the accumulated data from LOFAR we concentrate on a distance of 100~m from the shower axis since this is close to the distance where Cherenkov effects (relativistic time compression) are large and thus the pulse will have a flat frequency spectrum within our observing window.
From the maximum values at 100~m, as can be read from \figref{Stokes-ratio}, where $\phi=\pm 90^\circ$, one obtains $V/U\approx 1/3$ giving $\eta\approx 0.3$ using \eqref{Stokes-anal}.

In \figref{Stokes-V/U} the measured values for $U/I$ and $V/I$ are given for all antennas at a distance between 90 and 110~m from the core for the 114 high-quality events measured at LOFAR as given in Ref.~\cite{Bui16}. To restrict the analysis to antennas at an angle close to $90^\circ$ with respect to the $\vB$ axis, the additional condition $|\cos{\phi}|<0.5$ was imposed. A quality cut is applied where only those data are retained for which the measurement error in both $U/I$ and $V/I$ is smaller than 10\%.  This leaves us with 106 antenna readings. The average of the data given in \figref{Stokes-V/U} is $V/U=0.32$ giving $\eta\approx 0.31$ with a considerable spread as can be seen from the figure. This value supports the result derived from the single event shown in \figref{Stokes-ratio}.
The Stokes parameters are measured in the frequency band 30-80~MHz. Taking the central frequency as reference one obtains a time delay for the charge excess signal of approximately $\Delta t = 1$~ns using \eqref{eta}.

\begin{figure}
  \includegraphics[width=.4\textwidth]{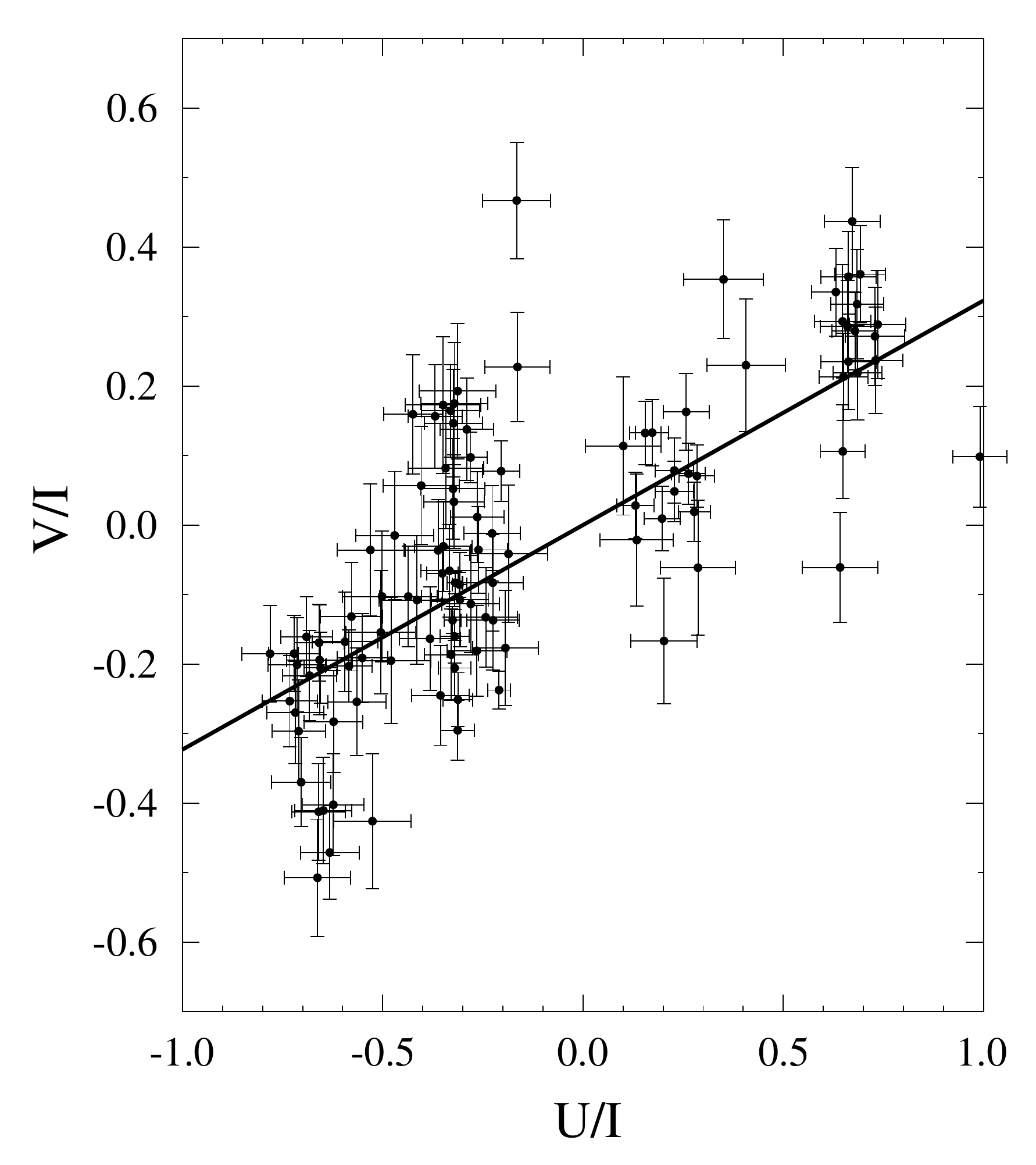}
  \caption{Values for Stokes parameters $V/I$ and $U/I$ at a distance between 90 and 110~m from the core for the high-quality events given in Ref~\cite{Bui16} after applying certain quality cuts (see text).
  \figlab{Stokes-V/U}}
\end{figure}

\section{Interpretation}

To understand the difference in the timing of the radio pulse emitted through the two mechanisms requires more subtle arguments.  Taking the $z$-axis along the shower and the $x$-axis along $\hat{e}_\vB$ it can be shown that the transverse current gives rise to the $x$-component of the vector potential,
\begin{equation}
A^\mu=j^\mu/D=j^x(\tret)/D
\end{equation}
where $j^\mu$ is the four current in the EAS and $D$ the retarded distance \cite{Krijn10}. The dependence on the height in the atmosphere, $z$, is expressed using the retarded time, $\tret=-z/c$.
The magnitude of the transverse current is roughly proportional to the number of particles in the shower, $N(\tret)$, thus $j^x(\tret)\propto N(\tret)$~\cite{Sch08}.
The charge excess in the EAS is due to the knock-out of electrons from air molecules and gives rise to the zeroth (time) and the $z$-components of the current. This excess charge is also roughly proportional to the number of particles in the EAS, $j^0(\tret)\propto N(\tret)$.
The height (or retarded time) dependence of the two currents is thus almost the same although, due to different dependencies on air pressure, the transverse current reaches its maximum at somewhat larger heights than the charge excess~\cite{Krijn10}.

The electric field is obtained from the vector potential as
\begin{equation}
\vec{E}= -\vec{\nabla}A^0 -d\vec{A}/dt \;.
\end{equation}
For the transverse current contribution this yields $\vec{E}= -d\vec{A}/dt$ (called magnetic emission for this reason and polarized in the $\hat{e}_\vB$ direction), while for the charge excess contribution this gives $\vec{E}= -dA^0/dr$ (thus called electric and polarized in the radial, $\hat{r}$ direction). Since $d\tret/dt\approx c(R/r)d\tret/dr$~\cite{Sch08} and $R$ is the distance to the emission point, this implies that for magnetic emission the lower parts of the shower (small $R$) are weighted less as compared to electric emission.
Numerical results show that the distance along the shower axis between the points at which the observed electric and magnetic emission reach their maxima is about 1~km, depending in detail on shower development and air slant depth profile.

An equivalent alternative explanation, in principle only valid in vacuum, is that the transverse current emission has a single lobe structure peaking at $\theta=0$, while the charge excess emission has a hollow cone structure, vanishing at $\theta=0$. The observed field strengths for both mechanisms thus depend on the viewing angle in a different way. Since the viewing angle changes as a function of time, the transverse current and charge excess emission will in general peak at different observer times, even when the magnitude of the current and the charge excess reach their maximum at the same time/location.

The first arrival of the radio signal at an antenna will be the same for the electric and magnetic emission mechanisms. However, due to the difference in emission-height dependence the time-structure of the pulse differs for the two mechanisms, where the electric pulse reaches its maximum at a later time. This shift is of the order of 1~km for a shower \Xmax\ at a height of 5~km. Ignoring Cherenkov effects, the arrival time depends on height ($h$) approximately as $c t=d^2/2h$ where $d$ is the distance from the shower axis. For $d=100$~m the arrival time difference is thus estimated to be 1~ns in agreement with the data. The  observed time delay depends on the observing frequency band due to a difference in structure of the two contributions as well as due to Cherenkov compression effects.

\section{Conclusions \& Summary}
We have shown a detailed comparison of the full set of Stokes parameters as function of distance to the shower axis for air showers measured with LOFAR. This is the first in-depth discussion of circular polarization in air showers and how it varies in accordance with expectations. The circular polarization is in good agreement with the predictions from current microscopic air shower simulations. In a macroscopic approach the circular polarization can be explained in terms of different emission mechanisms.
Due to different effective emission heights there is a time delay between the pulses generated by electric and magnetic emission resulting in a circular polarization component in the radio emission from an EAS.
This adds further justification to the present arguments for distinguishing two emission mechanisms based on their difference in the linear polarization of the signal~\cite{Sche14} also giving rise to the typical bean-shaped intensity profile seen in Ref.~\cite{Nel14b}.

The fact that we have a detailed understanding of the  {full set of} Stokes parameters for fair-weather events opens the possibility to use these in an analysis as presented in Ref.~\cite{Sche14} to extract atmospheric electric fields. {By being able to account for the circular component in addition to the linear components of the polarization will allow us to not only track the overall fields but study changes of the orientation of electric fields in more detail.}

\begin{acknowledgements}
The LOFAR cosmic ray key science project acknowledges funding from an Advanced Grant of the European Research Council (FP/2007-2013) / ERC Grant Agreement nr. 227610. The project has also received funding from the European Research Council (ERC) under the European Union's Horizon 2020 research and innovation programme (grant agreement No 640130).   We furthermore acknowledge financial support from FOM, (FOM-project 12PR304) and the Flemish foundation of scientific research (FWO-12L3715N). AN is supported by the DFG (research fellowship NE 2031/1-1).

LOFAR, designed and constructed by ASTRON, has facilities in several countries, that are owned by various parties (each with their own funding sources), and that are collectively operated by the International LOFAR Telescope foundation under a joint scientific policy.
\end{acknowledgements}

\appendix \section{Interpretation of $W$}\applab{W}

Most often Stokes parameters are used to express the polarization of a narrow-bandwidth (single-frequency) signal. In such a case one obtains $W=0$. For this case the result is independent of the number of time-samples, $n$ in \eqref{Stokes-anal}, that is used for the calculation.
For the case of our paper the signal has a large bandwidth which is reflected by a signal that has a narrow structure in time. Only when the complex signal in the two polarization directions differs by a constant phase one obtains $W=0$, independent of the number of time samples in \eqref{Stokes-anal}, while in the general cases one obtains $W/I>0$. For the general case the value of $W/I$ depends on $n$.
This can be visualized most clearly when the pulses in the two polarization directions, say $x$ and $y$, are shifted in time much more than their width. One then obtains $I\gg Q$ while the cross product of the two signals, $U+i V$, obviously almost vanishes, resulting in $W/I\approx 1$ i.e.\ exceeding zero by a sizable amount.
When the displacement of the pulses in the two polarization directions is much smaller than the intrinsic width of the pulse one has the situation that is pertinent to the case discussed in this paper. One then obtains a large value for $|U+i V|$ provided that the signals in the two polarization directions are commensurate. Again it is easy to see that for this case one obtains $W/I\approx 0$. Note that in this ratio the number of samples, $n$, drops out.

For real data (in contrast to a model calculation) the actual value of $W/I$ will be dominated by the noise level. For pure noise an average value is obtained of $W/I=\sqrt{1-1/n}$  where the $n$ dependence is explicit. Still, for relatively small values of $W/I$, the complex angle of $U+i V$ can be used to determine the time-lag, provided of course that $|U+i V|/I$ is sizable. This is shown in \figref{Stokes-V/U}.

\end{document}